\documentclass[11pt] {article}
\usepackage{epsfig}
\catcode`@=11

\def\@citex[#1]#2{\if@filesw\immediate\write\@auxout{\string\citation{#2}}\fi
  \def\@citea{}\@cite{\@for\@citeb:=#2\do
    {\@citea\def\@citea{,\penalty\@m}\@ifundefined
      {b@\@citeb}{{\bf ?}\@warning
       {Citation `\@citeb' on page \thepage \space undefined}}%
\hbox{\csname b@\@citeb\endcsname}}}{#1}}

\def\citer{\@ifnextchar [{\@tempswatrue\@citexr}{\@tempswafalse\@citexr[]}}

\def\@citexr[#1]#2{\if@filesw\immediate\write\@auxout{\string\citation{#2}}\fi
  \def\@citea{}\@cite{\@for\@citeb:=#2\do
    {\@citea\def\@citea{--\penalty\@m}\@ifundefined
       {b@\@citeb}{{\bf ?}\@warning
       {Citation `\@citeb' on page \thepage \space undefined}}%
\hbox{\csname b@\@citeb\endcsname}}}{#1}}
\catcode`@=12
\def\s{\hat{s}}
\def\mc{\hat{m}_c}
\def\bxsll{$B \rightarrow X_s \ell^+ \ell^- $}
\def\bxqll{$B \rightarrow X_q \ell^+ \ell^- $}

\def\absvcb{\left| V_{cb} \right|}
\title{  {\bf A Theoretical Reappraisal of Branching Ratios and CP
Asymmetries in the Decays $B \to (X_d,X_s) \ell^+ \ell^-$ 
and Determination of the CKM Parameters}}
\author{\vspace{1cm}\\
   {\bf A.~Ali${}^a$\thanks{E-mail address: ali@x4u2.desy.de} ~and 
G.~Hiller${}^b$ \thanks{E-mail address: Gudrun.Hiller@lnf.infn.it}}\\
        ${}^a$Deutsches Elektronen-Synchrotron DESY, Hamburg \\
   ${}^b$INFN, Laboratori Nazionali di Frascati,
                I-00044 Frascati, Italy
        \vspace{5mm}\\}
\date{}

\begin{document}
\setlength{\baselineskip}{5mm}

\maketitle
\begin{picture}(0,0)
       \put(325,275){DESY 98-187}
       \put(325,260){LNF-98/041(P)}  
       \put(325,245){December 1998}
\end{picture}
\vspace{-24pt}
\setlength{\baselineskip}{6mm}

\begin{abstract}
 We present a theoretical reappraisal of the branching ratios
and CP asymmetries for the decays $ B \to X_q \ell^+ \ell^-$, with 
$q=d,s$, taking into account current theoretical uncertainties in the 
description of the inclusive 
decay amplitudes from the long-distance contributions, an improved 
treatment of the renormalization scale dependence, and other parametric 
dependencies. Concentrating on the
partial branching ratios $\Delta {\cal B}(B \to X_q 
\ell^+ \ell^-)$, integrated over the invariant dilepton mass region
$1~\mbox{GeV}^2 \leq s \leq 6~\mbox{GeV}^2$, we
calculate theoretical precision on the charge-conjugate 
averaged partial  branching ratios $\langle \Delta{\cal B}_q \rangle=
(\Delta {\cal B}(B \to X_q \ell^+ \ell^-) + \Delta {\cal B}(\bar{B} \to 
\bar{X}_q \ell^+ \ell^-))/2$,
CP asymmetries in partial decay rates $(a_{CP})_q=(\Delta {\cal B}(B \to X_q 
\ell^+ \ell^-) -
\Delta {\cal B}(\bar{B} \to \bar{X}_q \ell^+ \ell^-))/(2 \langle \Delta{\cal 
B}_q \rangle)$, and the ratio of the branching ratios
$\Delta {\cal R} = \langle \Delta{\cal B}_d \rangle/\langle 
\Delta{\cal B}_s \rangle$. For the central values of the CKM parameters,
we find
$\langle \Delta {\cal B}_s \rangle =(2.22^{+0.29}_{-0.30}) \times 10^{-6}$,
$\langle \Delta {\cal B}_d \rangle =(9.61^{+1.32}_{-1.47}) \times 10^{-8}$,
$(a_{CP})_s =-(0.19^{+0.17}_{-0.19})\%$, $(a_{CP})_d 
=(4.40^{+3.87}_{-4.46})\%$, and $\Delta {\cal R} =(4.32 \pm 0.03)\%$. 
The dependence of $\langle \Delta{\cal B}_d \rangle$ and
$\Delta {\cal R}$ on the CKM parameters is worked out and the resulting
constraints on the unitarity triangle from an  eventual measurement of 
$\Delta {\cal R}$ are illustrated. 

\vspace*{1.5cm}
\centerline{(Submitted to Physics Letters B)}

\end{abstract}

\thispagestyle{empty}
\newpage
\setcounter{page}{1}

\section{Introduction}
  With the advent of new and upgraded experimental facilities in the
next year(s),
 flavour 
physics involving $B$ decays will come under minute experimental and 
theoretical scrutiny. The overriding interest in these experiments  
is in measuring  CP-violating asymmetries in partial $B$-decay rates, 
which will allow to  quantitatively test the Kobayashi-Maskawa \cite{CKM} 
paradigm of CP violation. In addition, 
the large number of $B$ hadrons anticipated to be produced at these 
facilities (estimated 
to be $O(10^8)$ - $O(10^{12})$) will allow to measure a number of  
flavour-changing-neutral-current (FCNC) processes involving the 
transitions $b \to sX$ and $b \to dX$, with $X=\gamma, g, \ell^+ \ell^-, 
\nu\bar{\nu}$, and $B^0$ - $\overline{B^0}$ mixings. In the context of the 
Standard Model (SM),  FCNC decays and mixings  measure the 
Cabibbo-Kobayashi-Maskawa (CKM) \cite{CKM} 
matrix elements, in particular $V_{td}$, $V_{ts}$ and $V_{tb}$. These
quantities can, in principle, also be measured directly in top quark 
decays $t \to q_i W^+$, with $q_i=d,s,b$.
A comparison of these matrix elements in the FCNC processes and direct
measurements in $t$ decays would provide one of the best strategies to 
search for new physics in $B$ decays. So far, only $V_{tb}$ has been 
directly measured at Fermilab, yielding $\vert V_{tb}\vert=0.99 \pm 0.15$ 
\cite{CDFVtb}.
 
  Present knowledge of $V_{td}$ owes itself to
the measurements of $\Delta M_d$, the mass difference in the $B^0$ -     
$\overline{B^0}$ complex.
 With the current world average $\Delta M_d=0.471 \pm 0.016$ (ps)$^{-1}$, 
the error on $V_{td}$ is dominated by theoretical uncertainty 
on the hadronic matrix element $f_{B_d}\sqrt{B_{B_d}}$, for which 
present Lattice-QCD estimates are $f_{B_d}\sqrt{B_{B_d}}= 215 \pm 35$ 
MeV \cite{Draper98}, yielding $0.0065 \leq \vert V_{td}V_{tb}^* \vert \leq 
0.010$. We also mention that a single event for
the charged kaon decay mode $K^+ \to \pi^+ \nu \bar{\nu}$ reported by the
Brookhaven E787 experiment, yielding ${\cal B}(K^+ \to \pi^+ \nu
\bar{\nu})= (4.2^{+9.7}_{-3.5}) \times 10^{-10}$, allows one to infer
$0.006 \leq \vert V_{td}V_{tb}^* \vert \leq 0.06$ \cite{E787}.
The branching ratio for the decay $B \to
X_s \gamma$ has led to a determination of the matrix element $V_{ts}$ 
\cite{Ali97}, yielding $\vert V_{ts} V_{tb}^*
\vert=0.0035 \pm 0.004$, with the error
dominated by the experimental error on the
branching ratio ${\cal B}(B \to X_s + \gamma)$ \cite{CLEO98,ALEPH98}.
These numbers can be taken as
the measurements of $\vert V_{td}\vert$ and $\vert V_{ts}\vert$ by 
assuming the value $V_{tb} \simeq 1$ from the
CKM unitarity, which holds to a very high accuracy \cite{PDG98}. 

  In this paper, we pursue the idea of measuring the FCNC semileptonic
decays $B \to X_s \ell^+ \ell^-$ and $B \to X_d \ell^+ \ell^-$, below 
the $J/\psi$- and above the $\rho,\omega$-resonance regions in the dilepton 
invariant mass, to determine $\vert V_{ts}\vert$ and $\vert V_{td}\vert$,
respectively, and  
 the ratio $\vert V_{td}/V_{ts}\vert$ from the ratio of the 
branching ratios. In this context, these decays and the related ones, 
$B \to X_s \nu \bar{\nu}$ and $B \to X_d \nu \bar{\nu}$,
were discussed some time ago \cite{AGM93}. The decays $B \to (X_s,X_d) 
\nu\bar{\nu}$ are practically free of long-distance 
complications \cite{buchallaisidorirey} and the renormalization-scale
dependence of the decay rates  has also been brought under control
\cite{Burasetal93}. Hence, these decays are theoretically remarkably clean
 but, unfortunately, they 
are difficult to measure in $\Upsilon(4S)$ decays and
out of question in hadronic collisions. Using the missing energy 
technique and LEP I data, the ALEPH collaboration 
has searched for the decays $B \to X_s \nu \bar{\nu}$ setting an upper bound 
${\cal B}(B \to X_s \nu \bar{\nu}) < 7.7 \times 10^{-4}$ (at 90\% C.L.)
\cite{ALEPHbsnunu},
which is a factor 20 away from the SM expectations \cite{Burasetal93}.
While the discovery of these decays looks formidable elsewhere,
a high luminosity $Z^0$-factory - being discussed in conjunction with an
$e^+ e^-$ linear collider \cite{Zerwas98}- looks like having the best chance
of measuring them. This possibility deserves a dedicated study.

 The possibility of determining  
$\vert V_{td}/V_{ts}\vert$ from the ratio of the invariant mass decay 
distributions
$\frac{dR}{ds} \equiv \frac{d{\cal B}}{ds}[B \to X_d \ell^+ \ell^-]/
 \frac{d{\cal B}}{ds}[B \to X_s \ell^+ \ell^-]$ away from the resonances
was revisited by Kim, Morozumi and Sanda \cite{KMS}. These authors
included the effects of the
leading order power corrections (in $1/m_b^2$) in the short-distance part 
of the dilepton invariant mass distribution and 
the long-distance contributions from the $c\bar{c}$-resonances, 
calculated in  Ref.~\cite{AHHM97}.
(For earlier-vintage derivations without the power corrections, 
see \cite{cslim89,amm91}.) We reanalyze the decays 
$B \to X_s \ell^+ \ell^-$ and $B \to X_d \ell^+ \ell^-$ and the ratio
of the branching ratios
$\Delta {\cal R} \equiv \int ds \frac{ d{\cal B}}{ds}[B \to X_d \ell^+ \ell^-]/
 \int ds \frac{ d{\cal B}}{ds}[B \to X_s \ell^+ \ell^-]$, integrated over a
kinematic range $q^2_{min} \leq s \leq q^2_{max}$, designed to minimize
the resonant contribution. Our theoretical treatment 
differs from that of Ref.~\cite{KMS} in a number of ways, 
summarized below.
\begin{itemize}
\item The dilepton invariant mass distributions in $B \to (X_s,X_d) 
\ell^+ \ell^-$ can be calculated in the context of the heavy quark 
effective theory (HQET) as a power expansion in regions far from
the resonances, thresholds and end-points \cite{AHHM97,buchallaisidorirey}.
Away from the $J/\psi, \psi^\prime,...$-resonances, the $1/m_c^2$-expansion 
provides, in principle, a
viable description of the non-perturbative contributions arising
from the $c\bar{c}$-loop \cite{buchallaisidorirey}. The 
contribution of the light quark $q\bar{q}$-loops, which is not 
CKM-suppressed 
in the decay $B \to X_d \ell^+ \ell^-$, can likewise be calculated 
by doing an expansion of the decay amplitudes 
in $\Lambda_{QCD}^2/q^2$ in regions of the dilepton squared mass satisfying
$q^2 \gg \Lambda_{QCD}^2$. Thus, the HQET framework provides an evaluation
of the invariant dilepton mass spectrum in these processes with the present
precision limited to the leading power corrections in $1/m_b^2$, $1/m_c^2$ 
and $\Lambda_{QCD}^2/q^2$. We present HQET-based calculations of
the decay rates, CP asymmetries and the ratio $\Delta {\cal R}$.
\item Away from the resonances and the end-points, the power corrections
in $1/m_b^2$ calculated in HQET and
in explicit wave function models, such as the Fermi motion (FM) model
\cite{aliqcd},
yield very similar invariant dilepton mass \cite{AHHM97} and hadron energy 
distributions \cite{AH98-12} in the decays $B \to X_q \ell^+ \ell^-$.
However, it is known that there are marked differences in estimates of the 
non-perturbative $c\bar{c}$-contribution, obtained by using the
$1/m_c^2$-corrections in the HQET
approach and alternative methods based on the  
Breit-Wigner-shaped resonant amplitudes \cite{AH98-3,MKS98}. Data may
eventually provide a discrimination against some of these approaches, but
currently at least four different variations on this
theme  exist 
in the literature \cite{buchallaisidorirey,AHHM97,KS96,LSW97}.
 This LD-uncertainty therefore 
compromises theoretical precision on decay rates and has to be taken into
account. We
calculate the theoretical 
uncertainties on the branching ratios for the decays $B \to (X_d,X_s)
\ell^+ 
\ell^-$, CP asymmetries and the ratio $\Delta {\cal R}$, showing 
numerically their 
impact on the determination of $\vert V_{ts} \vert$, $\vert V_{td} \vert$
and the CKM-Wolfenstein parameters $\rho$ and $\eta$
\cite{Wolfenstein} from an eventual measurement of these decays.
\item We reanalyze the renormalization scale dependence in the branching 
ratios for the decays
$B \to X_s \ell^+ \ell^-$ and $B \to X_d \ell^+ \ell^-$, using the method
employed by Kagan and Neubert in the 
radiative decay $B \to X_s + \gamma$ \cite{KN98}. This approach
avoids accidental cancellations among the individual
scale-dependent contributions but gives a 
larger scale ($\mu$)-dependence of the branching ratios than the method
of evaluating the same in the total branching ratio \cite{KMS}. The 
former is probably a more realistic estimate of the neglected higher order
corrections.
\end{itemize}

 We find that the partial branching ratio in the SM is
uncertain by typically
$\pm 13\%$ ($\pm 15\%)$ for the decay $B \to X_s \ell^+ \ell^-$
($B \to X_d \ell^+ \ell^-$), but the ratio $\Delta {\cal R}$ is
remarkably stable with typical error less than several percent. Hence,
$\Delta {\cal R}$ is well-suited to determine the ratio
$\vert V_{td}/V_{ts} \vert$. However,
 the scale-dependence of the CP asymmetries in $B \to
(X_s,X_d) \ell^+ \ell^-$ is found to be huge, reflecting the
(present) leading logarithmic
theoretical accuracy of the CP-odd parts of the amplitudes. 
Without the power corrections and fixing the scale to
$\mu=m_b$, the CP asymmetries in question  
have been studied earlier in Ref.~\cite{KSCP97}. 
We point out that these estimates are uncertain by
almost $\pm 100\%$ due to the sensitive scale-dependence 
and their stabilization requires next-to-leading
order corrections. In
the case of the CP-even parts,
we recall that the inclusion of the explicit $O(\alpha_s)$
corrections in the matrix elements has reduced the scale dependence
of the decay rates considerably \cite{burasmuenz,misiakE}.
 

This paper is organized as follows: In section 2, we briefly review the
derivation of the matrix elements and dilepton invariant mass distributions
for the decays $B \to (X_s,X_d) \ell^+ \ell^-$ including long-distance
contributions in the four approaches: (i) AMM \cite{amm91,AHHM97}, (ii) KS 
\cite{KS96}, (iii) LSW \cite{LSW97} and (iv) HQET \cite{buchallaisidorirey}. 
The partially integrated 
branching ratios and CP asymmetries are presented in section 3 where we
also specify our input parameters. We show the scale dependence of the
branching ratios $\Delta {\cal B}(\bar{B} \to \bar{X}_s \ell^+ \ell^-)$
and $\Delta {\cal B}(\bar{B} \to \bar{X}_d \ell^+ \ell^-)$  in the AMM
approach and the contributions arising from the individual Wilson 
coefficients. 
We also present a comparative numerical study of the quantities
$\langle \Delta {\cal B}_s \rangle$, $\langle \Delta {\cal B}_d \rangle$,
$(a_{CP})_s$ and  $(a_{CP})_d$ in the four mentioned approaches.
Uncertainties arising from the other parameters ($m_b$, $m_t$ and 
$\Lambda_{QCD}^{(5)}$) are worked out numerically.
With this we calculate the overall theoretical errors in these quantities
and the ratio $\Delta {\cal R}$ and their impact on the determination of
the CKM parameters.
 Finally, section 4 contains a  brief comparison of 
the theoretical precision  of $\vert 
V_{td}/V_{ts}\vert$ in the decays $B \to (X_s,X_d) \ell^+ \ell^-$ with
that of other methods proposed in the literature  
to determine the same ratio.

\section{$B \to (X_d,X_s) \ell^+ \ell^-$ Decays in the Effective
Hamiltonian Approach}

We work in the effective Hamiltonian approach, which is based on integrating
out the heavy degrees of freedom $(t,W^\pm,Z^0)$, in the SM.
The resulting effective Hamiltonian  for the decays $B \to (X_d,X_s) 
\ell^+ \ell^-$, ${\cal H}_{eff}(b \to q \ell^+ \ell^-)$, can be expressed as
follows: 
\begin{eqnarray} {\cal H}_{eff} (b \to q \ell^+ \ell^-)
  = -\frac{4 G_F}{\sqrt{2}} V_{tq}^*V_{tb} \sum_{i=1}^{10} C_i O_i  
+\frac{4 G_F}{\sqrt{2}} V_{uq}^* V_{ub}
\left[ C_1 ({O_1}^{(u)}-O_1) + C_2 ({O_2}^{(u)}-O_2) \right] \; ,
\end{eqnarray}
where $V_{ij}$ are the CKM matrix elements.
The $C_i$ are the Wilson coefficients,
which depend, in general, on the renormalization scale $\mu$,
except for $C_{10}$, and can be seen in leading logarithmic approximation
in \cite{burasmuenz}. The operators are defined as follows:
\begin{eqnarray}
 O_1 &=& (\bar{q}_{L \alpha} \gamma_\mu b_{L \alpha})
               (\bar{c}_{L \beta} \gamma^\mu c_{L \beta})\; ,
 \nonumber \\
 O_2 &=& (\bar{q}_{L \alpha} \gamma_\mu b_{L \beta})
               (\bar{c}_{L \beta} \gamma^\mu c_{L \alpha})\; ,
 \nonumber \\
 O_3 &=& (\bar{q}_{L \alpha} \gamma_\mu b_{L \alpha})
               \sum_{q'=u,d,s,c,b}
               (\bar{q}^{\prime}_{L \beta} \gamma^\mu q'_{L \beta})\; ,
 \nonumber \\
 O_4 &=& (\bar{q}_{L \alpha} \gamma_\mu b_{L \beta})
                \sum_{q'=u,d,s,c,b}
               (\bar{q}^{\prime}_{L \beta} \gamma^\mu q'_{L \alpha})\; ,    
 \nonumber \\
O_5 &=& (\bar{q}_{L \alpha} \gamma_\mu b_{L \alpha})
               \sum_{q'=u,d,s,c,b}
               (\bar{q}^{\prime}_{R \beta} \gamma^\mu q'_{R \beta})\; ,    
 \nonumber \\
O_6 &=& (\bar{q}_{L \alpha} \gamma_\mu b_{L \beta})
                \sum_{q'=u,d,s,c,b}
               (\bar{q}^{\prime}_{R \beta} \gamma^\mu q'_{R \alpha})\; ,      
 \nonumber \\
O_7 &=& \frac{e}{16 \pi^2}
 \bar{q}_{\alpha} \sigma_{\mu \nu} (m_b R + m_q L) b_{\alpha}
                F^{\mu \nu}\; ,
                                     \nonumber \\
 O_8 &=& \frac{g}{16 \pi^2}
    \bar{q}_{\alpha} T_{\alpha \beta}^a 
                \sigma_{\mu \nu} (m_b R + m_q L)  
          b_{\beta} G^{a \mu \nu},  \nonumber \\
 O_9 &=& \frac{e^2}{16 \pi^2} \bar{q}_\alpha 
             \gamma^{\mu} L b_\alpha
\bar{\ell} \gamma_{\mu} \ell \; , \nonumber\\
 O_{10} &=& \frac{e^2}{16 \pi^2} \bar{q}_\alpha \gamma^{\mu} L
b_\alpha \bar{\ell} \gamma_{\mu}\gamma_5 \ell \; ,
\end{eqnarray}
where $L$ and $R$ denote chiral projections, $L(R)=1/2(1\mp \gamma_5)$.
Here, unitarity of the CKM matrix has been used in writing the flavour 
structure 
of a generic FCNC $b \to q$ transition amplitude ${\cal{T}}^{(q)}$ in
the form
\begin{eqnarray}
{\cal{T}}^{(q)}=\sum_{i=u,c,t} \lambda_i^{(q)}{\cal{T}}_i
=\lambda_t^{(q)}({\cal{T}}_t-{\cal{T}}_c)+
\lambda_u^{(q)}({\cal{T}}_u-{\cal{T}}_c) \; ,
\label{eq:T}
\end{eqnarray}
where $\lambda_i^{(q)}=V_{iq}^* V_{ib}$ and $q=d,s$.
For the $b \to s$ transitions, the second term in Eq.~(\ref{eq:T}) can be 
safely neglected as $\lambda_u^{(s)} \ll \lambda_t^{(s)}$. However,
for the $b \to d$ transitions, the CKM factors $\lambda_u^{(d)}$ and 
$\lambda_t^{(d)}$ are of the same order and hence all terms 
in Eq.~(\ref{eq:T}) must be 
kept. The operator basis given in Eq.~(1) has been written 
in accordance with Eq.~(\ref{eq:T}) and includes the 
Four-Fermi operators containing a $u\bar{u}$ pair,
\begin{eqnarray}
 {O_1}^{(u)} &=&  (\bar{q}_{L \alpha} \gamma_\mu b_{L \alpha})
               (\bar{u}_{L \beta} \gamma^\mu u_{L \beta})\; ,
\nonumber \\
 {O_2}^{(u)} &=&  (\bar{q}_{L \alpha} \gamma_\mu b_{L \beta})
               (\bar{u}_{L \beta} \gamma^\mu u_{L \alpha}) \; .
  \end{eqnarray}

The matrix element for the decays 
$ b \to q \ell^+ \ell^- \; (q=d,s)$ can be written as 
\begin{eqnarray} 
        {\cal{M}} (b \to q \ell^+ \ell^-) & = & 
     \frac{G_F \alpha}{\sqrt{2} \pi} \, V_{tq}^\ast V_{tb} \, 
        \left[ \left( C_{9q}^{\mbox{eff}} - C_{10} \right) 
                \left( \bar{q} \, \gamma_\mu \, L \, b \right)
                \left( \bar{\ell} \, \gamma^\mu \, L \, \ell \right) 
                + \left( C_{9q}^{\mbox{eff}} + C_{10} \right) 
                \left( \bar{q} \, \gamma_\mu \, L \, b \right)
                \left( \bar{\ell} \, \gamma^\mu \, R \, \ell \right)  
                \right. \nonumber \\
        &-& \left. 
  2 C_7^{\mbox{eff}}\left(\bar{q} \, i \, \sigma_{\mu \nu} \, 
        \frac{q^\nu}{q^2} (m_q L + m_b R) \, b \right)  
                \left( \bar{\ell} \, \gamma^\mu \, \ell \right) 
                \right] \; .
\end{eqnarray} 
Here $q^{\nu} \equiv p_+^{\nu}+p_{-}^{\nu}$ denotes the Four-momentum of the
invariant dilepton system, where $p_{\pm}$ are the corresponding momenta of 
the $\ell^{\pm}$; $s\equiv q^2$ is the invariant dilepton mass squared. 
The effective coefficients of $O_9$ are given by
\begin{equation}
C_{9q}^{\mbox{eff}} (\hat{s}) = C_9 \eta(\s) + {Y}^q (\hat{s}) \; .
\label{eqn:c9eff}
\end{equation}
The functions $\eta(\s)$ and $Y^q(\s)$ represent the ${\cal{O}}(\alpha_s)$ 
correction \cite{jezkuhn}
and the (perturbative) one loop matrix element of the 
Four-Fermi operators \cite{burasmuenz,misiakE}, respectively. 
We have in the (naive dimensional regularization) NDR-scheme, which we use 
throughout our work,
\begin{eqnarray} 
        {Y}^q (\s) & = & g(\mc,\s)
                \left(3 \, C_1 + C_2 + 3 \, C_3
                + C_4 + 3 \, C_5 + C_6 \right)
\nonumber \\
        &- &  \frac{1}{2} g(1,\s)
                \left( 4 \, C_3 + 4 \, C_4 + 3 \,
                C_5 + C_6 \right) 
         - \frac{1}{2} g(0,\s) \left( C_3 +   
                3 \, C_4 \right) \nonumber \\
        &+ &     \frac{2}{9} \left( 3 \, C_3 + C_4 +
                3 \, C_5 + C_6 \right)
          -{V_{uq}^* V_{ub} \over V_{tq}^* V_{tb}} (3 C_1 +C_2)
               (g(0 , \s)-g(\mc, \s)) \; , 
\label{eq:y}
\end{eqnarray}
where we have introduced the dimensionless variable $\s \equiv q^2/m_b^2$ and
$\hat{m}_c \equiv m_c/m_b$.
The functions $\eta(\s)$ and $g(z,\s)$ can be seen elsewhere 
 \cite{burasmuenz,AH98-3}. Note that the
renormalization scheme-dependence of the function $Y^q(\s)$
cancels with the corresponding one in $C_9$.
The effective coefficient of the $b s \gamma$ 
vertex is given by $C_7^{\mbox{eff}}=C_7-C_{5}/3-C_6$ \cite{effhamburas}.

The dilepton invariant mass spectrum including power corrections 
in the HQET approach in \bxqll decays can be written as:
\begin{eqnarray}
\frac{d {\cal{B}}}{d \s}=\frac{d {\cal{B}}^0}{d \s}
+\frac{d {\cal{B}}^{1/m_b^2}}{d \s}+
\frac{d {\cal{B}}^{1/q^2}}{d \s} \; ,
\end{eqnarray}
where the first term corresponds to the parton model 
\cite{burasmuenz,misiakE}, the second term
accounts for the ${\cal{O}}(1/m_b^2)$ power corrections \cite{AHHM97}, and
the last term accounts for the non-perturbative interaction of a virtual 
$u\bar{u}$- and $c\bar{c}$-quark loop with soft gluons. The explicit 
expression for $d {\cal{B}}^{1/q^2}/d\s$ for $m_q=0$ can be 
deduced from the literature \cite{buchallaisidorirey}
\begin{eqnarray}
\label{cupower}
\frac{d {\cal{B}}^{1/q^2}}{d \s} & \! \! \!= 
\! \! \!& -{\cal{B}}_0 C_2 \lambda_2 \frac{32}{27} (1-\s)^2  \\
& \! \! \! \times \! \! \!& Re \left\{ 
\left[ C_7^{\mbox{eff}*} \frac{(1+6 \s-\s^2)}{\s}+
  C_{9q}^{\mbox{eff}(0)*}(\s) (2+\s) \right]  
  \left[ \frac{F(s,m_c)}{m_c^2}-
\frac{\lambda_u^{(q)}}{\lambda_t^{(q)}}
(\frac{F(s,m_u)}{m_u^2}-\frac{F(s,m_c)}{m_c^2}) \right] \nonumber \right. \\ 
& \! \! \!+ \! \! \! & \left. 
\left[ (3 C_1 +C_2) (g(0 , \s)-g(\mc, \s)) \right]^* (2+\s)
\left[ |\frac{\lambda_u^{(q)}}{\lambda_t^{(q)}}|^2 
(\frac{F(s,m_u)}{m_u^2}-\frac{F(s,m_c)}{m_c^2}) -
\frac{\lambda_u^{(q)}}{\lambda_t^{(q)}}
\frac{F(s,m_c)}{m_c^2}\right]
\right\} \nonumber \; .
\end{eqnarray}
The branching ratio for \bxqll is expressed in terms
of the measured semileptonic branching ratio ${\cal B}_{sl}$
for the decays $B \to X_c \ell \nu_\ell$. This fixes
the normalization
\begin{equation} 
        {\cal B}_0 \equiv
                {\cal B}_{sl} \frac{3 \, \alpha^2}{16 \pi^2} \frac{
    {\vert V_{tq}^* V_{tb}\vert}^2}{\absvcb^2} \frac{1}{f(\mc) \kappa(\mc)}
                \; ,
\label{eqn:seminorm}
\end{equation}
where $f(\mc), \kappa(\mc)$ can be seen, for example, in \cite{AHHM97}.
The function $F(s,m) \equiv F(r)$ with 
$r=s/(4 m^2)$ is given in \cite{buchallaisidorirey}. In the region $r \gg 1$,
$F(s,m_u)/m_u^2 \propto 1/s $. The condition $r\gg 1$ is well satisfied, for 
example, for
$q^2 \geq 1.0 ~{\mbox{GeV}}^2$ (for which $r > 25$). In this region, the
operator 
product expansion (OPE) is not  in '$1/m_u^2$' but in 
$\Lambda_{QCD}^2/q^2$. Hence, there is a sufficiently large region in 
$q^2$ where the OPE holds in $1/m_b^2$, $1/m_c^2$ and $\Lambda_{QCD}^2/q^2$.
Note also that for the terms proportional to the power corrections, we use
$C_{9q}^{\mbox{eff}(0)*}(\s)$ which equals $C_{9q}^{\mbox{eff}*}(\s)$
with $\eta(\s)=1$.

In \bxqll decays $c\bar{c}$-resonances are present via
$B \to X_q+ (J/\psi,\psi^\prime,...)\to X_q \ell^+ \ell^-$. 
Their implementation and the corresponding uncertainties in the $B \to 
X_s\ell^+\ell^-$ case have been discussed recently by us 
\cite{AH98-3}. 
 There are at least four different Ans\"atze advocated in the
literature in this context, summarized below.
\begin{itemize}
\item The HQET-based approach \cite{buchallaisidorirey}, 
where the non-perturbative $c\bar{c}$-contribution
 away from the $(J/\psi,\psi^\prime,...)$-resonances is 
implemented by the $1/m_c^2$ terms in the expression for $d{\cal 
B}^{1/q^2}/d\s$. 
%
%
\item  One could add the resonant $c\bar{c}$-contribution,
 parametrized using a Breit-Wigner shape with the 
normalizations fixed by data, to the complete perturbative contribution
resulting from the $c\bar{c}$-loop. This
scheme has been used in a number of papers \cite{amm91,AHHM97,KMS,AH98-3}.
\end{itemize} 
The effective coefficients including the $c\bar{c}$-resonances
are defined as 
\begin{equation}
C_{9q}^{\mbox{eff}} (\hat{s}) \equiv  C_9 \eta(\s) + {Y}^q (\hat{s})
 + {Y_{res}}^q (\hat{s}) \; ,
\label{eqn:c9effres}
\end{equation}
where $Y^q(\s)$ has been given earlier and ${Y_{res}}^q(\s)$ in this 
scheme is defined as:
\begin{eqnarray}
         {Y_{res}}^q(\s) & = & 
                      \frac{3 \pi}{\alpha^2} \kappa \,
       \left( -{ V_{cq}^* V_{cb} \over V_{tq}^* V_{tb}}  C^{(0)}
              -{V_{uq}^* V_{ub} \over V_{tq}^* V_{tb}} ( 3 \, C_3
              + C_4 + 3 \, C_5 + C_6 )\right) \nonumber \\
        & & \times \sum_{V_i = \psi(1s),..., \psi(6s)}
      \frac{\Gamma(V_i \rightarrow \ell^+ \ell^-)\, M_{V_i}}{
      {M_{V_i}}^2 - \s \, {m_b}^2 - i M_{V_i} \Gamma_{V_i}} \; ,
                \label{eq:yres} 
\end{eqnarray}
with $C^{(0)} \equiv 3 C_1 + C_2 + 3 C_3 + C_4 + 3 C_5 + C_6$.
In what follows we shall neglect the part 
$\sim {V_{uq}^* V_{ub} \over V_{tq}^* V_{tb}}$ in Eq.~(\ref{eq:yres})
in our numerical analysis, since the particular combination of the Wilson 
coefficients appearing in this term  is strongly suppressed compared to  
$C^{(0)}$.
Further, since data only determines the product $\kappa C^{(0)}=0.875$
\cite{PDG98}, we keep this fixed. For ease of writing, we call this
approach the AMM approach \cite{amm91}.
 
The remaining two approaches are the following:
\begin{itemize}
\item The LSW-approach \cite{LSW97}: Here, for the non-resonant  
$c\bar{c}$-contribution, only the constant term in
$g(\hat{m}_c,\s)$ is kept. Calling it $\tilde{g}(\hat{m}_c,\s)$, it is
given by $\tilde{g}(\hat{m}_c,\s) = -\frac{8}{9} \ln(m_b/\mu) -\frac{8}{9}\ln
\hat{m}_c + \frac{8}{27}$. The resonant $c\bar{c}$ part is essentially as 
given in Eq.~(\ref{eq:yres}).
\item The KS-approach \cite{KS96}, in which the function 
$C_{9q}^{\mbox{eff}}(\s)$
is parametrized using a dispersion approach. 
For details and further discussions of this approach, we refer to 
\cite{KS96,AH98-3}.
 \end{itemize}
In $B \to X_d \ell^+ \ell^-$ decays, in addition to the $c\bar{c}$ bound 
states, also the  $u \bar{u}$ bound states 
have to be included in the decay amplitudes. We have calculated 
the dilepton invariant mass distribution, using the Breit-Wigner shape for 
the resonances, as discussed earlier, and
taking the widths and partial leptonic widths from
the Particle Data Group \cite{PDG98}.
  However, numerically the $u\bar{u}$-resonant part is less important, 
as the
leptonic branching ratios ${\cal B}(V^0 \to e^+ e^-)$ and ${\cal B}(V^0 \to 
\mu^+ \mu^-)$
for the dominant resonances $V^0=\rho^0, \omega$ are small \cite{PDG98}.
 Moreover, their effect is 
reduced by imposing a cut on the dilepton invariant mass, say $q^2 > 1 
\mbox{GeV}^2$, which we have explicitly checked. 
Higher states like $\rho^\prime, \omega^\prime$ have larger widths and are 
thus expected to play minor roles due to their smaller branching ratios
in dilepton pairs. 

In the three approaches discussed above (AMM,LSW,KS) we include the 
$1/m_b^2$-corrections, calculated in the phenomenological 
Fermi motion model (FM) \cite{aliqcd}, which implements such effects in 
terms of the  $B$-meson wave function effects. 
The implementation of the FM model in \bxsll decays in the dilepton 
invariant 
mass distribution can be seen in \cite{AHHM97}, which we also adopt here for
the calculations of the distributions in $B \to X_d \ell^+ \ell^-$. We note
that the branching ratios in the HQET-based $1/m_b^2$ approach and the 
FM-model are very close to each other for identical values of the input
parameters. 

\section{Branching Ratios and CP Asymmetries in \bxqll}
\subsection{Numerical input and definitions of the partial branching
ratios and CP asymmetries}
We now specify how we determine theoretical uncertainties in the
branching ratios, the ratio $\Delta {\cal R}$, and CP asymmetries in the 
decays $B \to (X_{s}, X_{d}) \ell^+ \ell^-$.
\begin{table}[h]
        \begin{center}
        \begin{tabular}{|l|l|}
        \hline
        $m_W$                   & $80.41$ GeV \\
        $m_Z$                   & $91.1867$ GeV \\
        $\sin^2 \theta_W $      & $0.2255$ \\
        $m_s$                   & $0.2$ GeV   \\
        $m_d$                   & $0.01 $ GeV   \\
        $m_b$                   & $4.8 \pm 0.2 $ GeV \\
        $m_t$                   & $173.8 \pm 5.0$ GeV     \\
        $\mu$                   & ${m_{b}}^{+m_{b}}_{-m_{b}/2}$        \\
        $\Lambda_{QCD}^{(5)}$   & $0.220^{+0.078}_{-0.063}$ GeV       \\
        $\alpha^{-1}$     & 129           \\
        $\alpha_s (m_Z) $       & $0.119 \pm 0.0058$ \\
        ${\cal B}_{sl}$         & $(10.4 \pm 0.4)$ \%   \\
        \hline
        \end{tabular}
        \end{center}
\caption{\it Default values of the input parameters and the $\pm 1~\sigma$ 
errors on the sensitive parameters used in our numerical calculations.}
\label{parameters}
\end{table}
The dispersion in the values of  
the obsevables due to the errors in the input parameters 
$m_b,\mu, m_t$, $\alpha_s(m_Z)$ (equivalently $\Lambda_{QCD}^{(5)}$),
and ${\cal B}_{sl}$, 
given in  Table~\ref{parameters}, is calculated by varying one parameter
at a time. 
To estimate the uncertainty from the $b$-quark mass in the FM 
model, we explore 
the parameter space of this model with three sets of parameters:
$(p_F,m_q)=(520,280),(450,0),(245,0)$ in (MeV,MeV), which correspond to an 
effective $b$-quark mass of 
$m_b^{\mbox{eff}}=4.6,4.8,5.0$ GeV, respectively. We set 
$m_c=m_b^{\mbox{eff}}(m_b)-3.4$ GeV in both the FM-model and HQET  analysis.
Comparison with the HQET prediction \cite{AHHM97} is worked out for
$\lambda_1=-0.20~\mbox{GeV}^2$  and  $\lambda_2=0.12~\mbox{GeV}^2$, as
the dependence of the branching ratios on these parameters is small.
The individual errors are then added in quadrature to get the final
cumulative error.%

We proceed by defining the partly integrated branching ratios 
($q=s,d)$:
 \begin{eqnarray}
\Delta {\cal{B}}_q \equiv \int_{q^2_{\rm min}}^{q^2_{\rm max}} dq^2 
\frac{d {\cal{B}}(B \to X_q \ell^+ \ell^-)}{dq^2} \; ,
\end{eqnarray}
together with $\Delta \bar{{\cal{B}}}_q$,  for the CP-conjugate decays
$\bar{B} \to \bar{X}_q \ell^+ \ell^-$, 
and the branching ratio averaged over the charge-conjugated states:
\begin{eqnarray}
\label{eq:delbr}
\langle \Delta {\cal{B}}_q \rangle \equiv 
\frac{\Delta {\cal{B}}_q+\Delta \bar{{\cal{B}}}_q}{2} \; ,
\end{eqnarray}
The CP asymmetry in the partial rates for $B \to X_q \ell^+ \ell^-$ is
defined as:
\begin{eqnarray}
\label{eq:acp}
(a_{CP})_q \equiv \frac{\Delta {\cal{B}}_q-\Delta \bar{{\cal{B}}}_q}
{\Delta {\cal{B}}_q+\Delta \bar{{\cal{B}}}_q} \; .
\end{eqnarray}
We further decompose the partial branching ratios $\Delta {\cal{B}}_q$ in 
terms of the CKM factors
 \begin{eqnarray}
\Delta {\cal{B}}_q =(|\lambda_t^{(q)}|^2 D_t^{(q)}+
|\lambda_u^{(q)}|^2 D_u^{(q)}+Re(\lambda_t^{(q)*} \lambda_u^{(q)}) D_r^{(q)} +
Im(\lambda_t^{(q)*} \lambda_u^{(q)}) D_i^{(q)})/|V_{cb}|^2 \; ,
\end{eqnarray}
from which the CP conjugated branching ratio $\Delta \bar{{\cal{B}}}_q$ 
can be 
obtained by substituting $\lambda_{u,t}^{(q)} \to \lambda_{u,t}^{(q) *}$. 
Hence, the charge-conjugate averaged branching ratio $\langle \Delta 
{\cal{B}}_q \rangle$ is obtained from 
$\Delta {\cal{B}}_q$ by dropping the $Im(\lambda_t^{(q)*} 
\lambda_u^{(q)})$ term. The CP asymmetry is given by the expression:
\begin{eqnarray}
\label{eq:deli}
(a_{CP})_q=Im(\lambda_t^{(q)*} \lambda_u^{(q)}) D_i^{(q)}/
(|V_{cb}|^2 \langle \Delta {\cal{B}}_q \rangle) \; .
\end{eqnarray}
The functions $D_j^{(q)}$, $j=t,u,r,i$ depend on the input parameters,
which we have specified in Table 1, and on the interval in $q^2$, specified
by $q^2_{min}$ and $q^2_{max}$. We shall work always above the  
($\rho$, $\omega$)- and below the  $J/\psi$-resonances
in the so-called low-$q^2$ region with $q^2_{min}$ and $q^2_{max}$
taken as
\begin{eqnarray}
\label{eq:cuts}
q^2_{\rm min}= 1.0 ~{\mbox{GeV}}^2 \leq q^2 \leq 6.0 ~{\mbox{GeV}}^2=
q^2_{\rm max} \; .
\end{eqnarray}

We use the Wolfenstein representation of the CKM matrix  
\cite{Wolfenstein} with $A=0.819$ and $\lambda= 0.2196$ fixed, 
as the errors on these quantities are small \cite{PDG98}.
The other two parameters $(\rho, \eta)$ are implicitly the subject of the 
present work.
Defining $\bar{\rho}=\rho (1-\frac{\lambda^2}{2})$ and
$\bar{\eta}=\eta (1-\frac{\lambda^2}{2})$, we have up to terms of 
order $\lambda^6$ \cite{BLO94}:
\begin{eqnarray}
\lambda_u^{(s)}=A \lambda^4 (\rho-i \eta) \; ,& &
\lambda_t^{(s)}=- A \lambda^2 
\left[ 1-\frac{\lambda^2}{2}+ \lambda^2 (\rho-i \eta) \right] \; ,\\
\lambda_u^{(d)}=A \lambda^3 (\bar{\rho}-i \bar{\eta})\; , & &
\lambda_t^{(d)}=A \lambda^3 (1-\bar{\rho}+i \bar{\eta}) \; ,
\end{eqnarray}
and $V_{cb}=A \lambda^2$.
It follows that $\left|{V_{td} \over V_{ts}}\right|^2=
\lambda^2 (1+\lambda^2 (1-2 \bar{\rho}))((1-\bar{\rho})^2+\bar{\eta}^2)
+ O(\lambda^6)$. Global fits of the CKM parameters have been performed
in a number of
papers \cite{smele98,parodi98,alilondon98}, with very similar (though 
not identical) results. 
For illustration, we shall use the results of the CKM fits
from Ref.~\cite{smele98}, yielding:
\begin{eqnarray}
\label{eq:ckmvalues}
\rho=0.155^{+0.115}_{-0.105} \; , & & \eta=0.383^{+0.063}_{-0.060} \; .
\end{eqnarray}

\subsection{Parametric dependence of the branching ratios and 
CP asymmetries}
\begin{figure}[t]
     \mbox{ }\hspace{-0.7cm}
     \begin{minipage}[t]{8.2cm}
     \mbox{ }\hfill\hspace{1cm}(a)\hfill\mbox{ }
     \epsfig{file=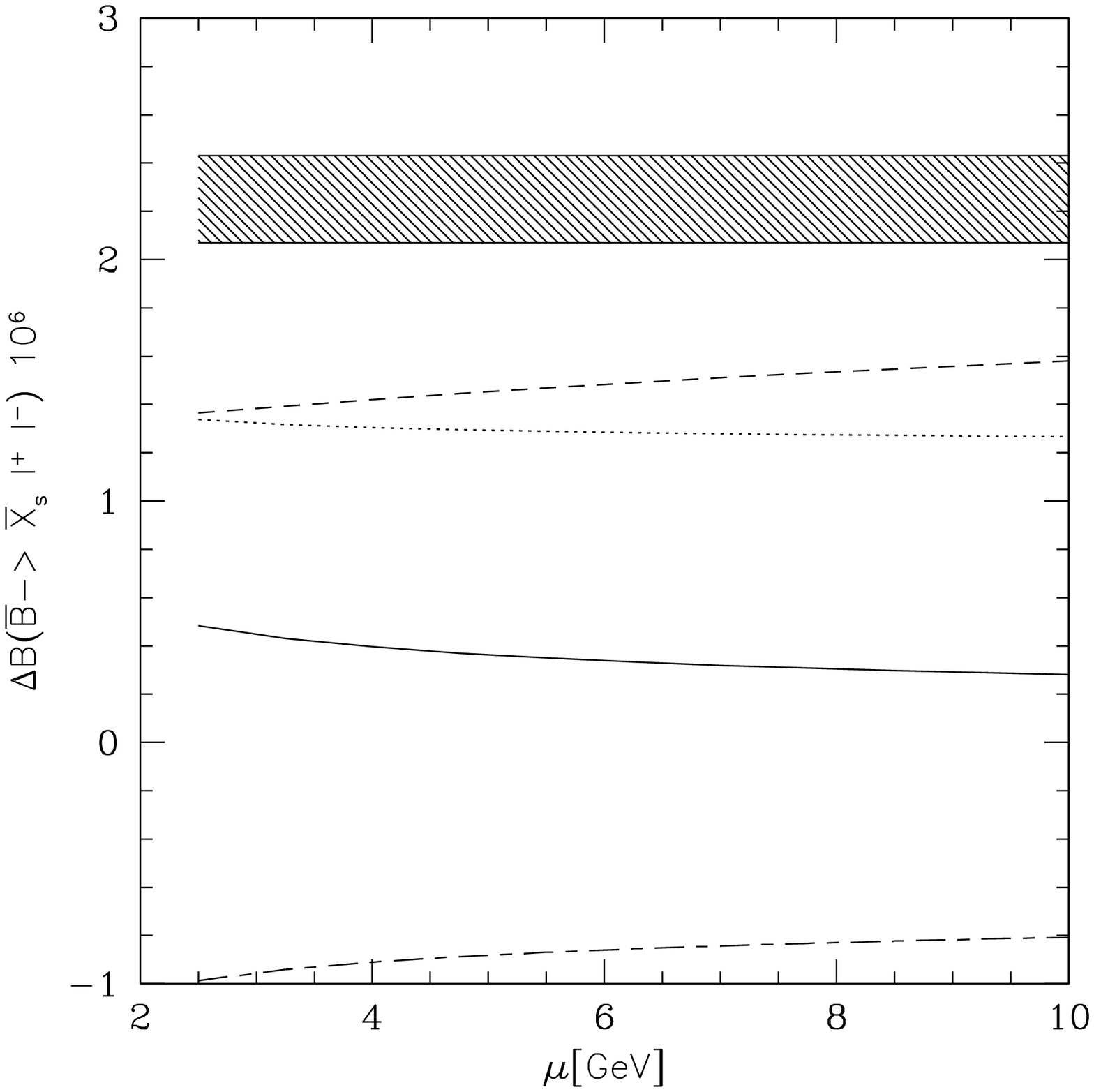,width=8.2cm}
     \end{minipage}
     \hspace{-0.4cm}
     \begin{minipage}[t]{8.2cm}
     \mbox{ }\hfill\hspace{1cm}(b)\hfill\mbox{ }
     \epsfig{file=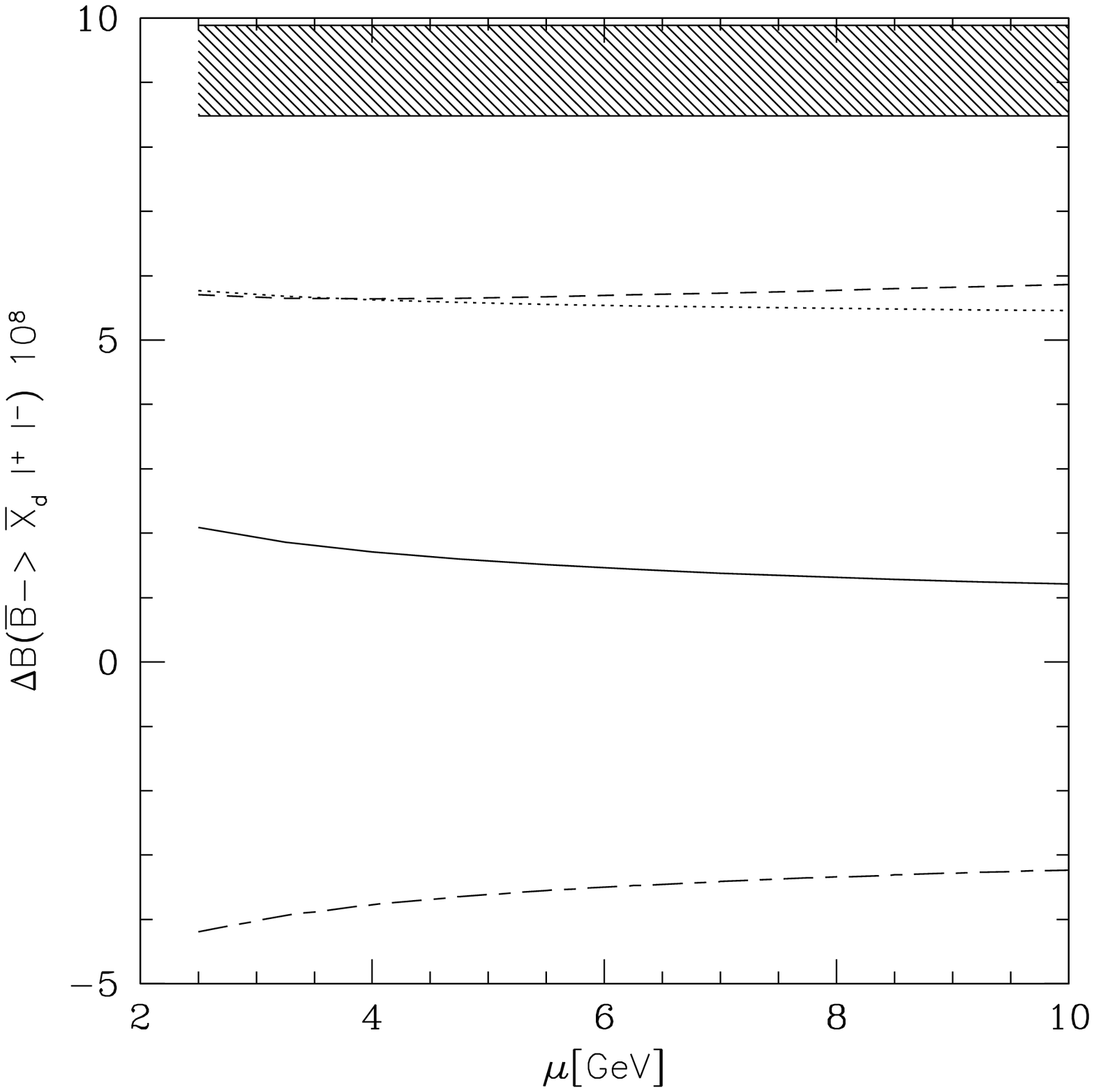,width=8.2cm}
     \end{minipage}
     \caption{\it
Renormalization scale $(\mu)$-dependence of the individual terms and the
 partly integrated branching 
ratios $\Delta B_s$ for  the decay $\bar{B} \to \bar{X}_s \ell^+ \ell^-$ 
(a) and  $\Delta B_d$ for $\bar{B} \to \bar{X}_d \ell^+ 
\ell^-$ (b), calculated in the AMM-approach.
The solid, dotted, dashed, long-short dashed curves correspond to the 
contributions proportional to the effective Wilson coefficients  
$|C_7^{\mbox{eff}}|^2,|C_{10}|^2,|C_9^{\mbox{eff}}|^2$ and
$Re(C_7^{\mbox{eff}}C_9^{\mbox{eff}})$, respectively. The resulting $\mu$
uncertainty in the branching ratio, obtained by adding the weighted 
errors in quadrature, is indicated by the shaded area.} \label{fig:mudep}
\end{figure}
We study the scale ($\mu$)-dependence of the branching ratios
along the lines followed in \cite{KN98} in the $B 
\to X_s \gamma$ case. Thus, instead of varying
the scale $\mu$ between $m_b/2$ and $2 m_b$ in the full expression for the
respective branching ratios (the naive method), the scale-dependence 
of the individual terms involving different
Wilson coefficient combinations is calculated independently and the 
resulting errors are added in quadrature.
It is a 
conservative approach and avoids the possibility of accidental cancellations
of the scale-dependence in the various terms, which takes place in the SM
in both the $B \to X_s \gamma$ case \cite{KN98} and in $B \to X_q \ell^+ 
\ell^-$, as shown here. For the branching ratio in \bxqll decays the relevant 
coefficients are:
$|C_{10}|^2,|C_9^{\mbox{eff}}|^2,Re(C_7^{\mbox{eff}}C_9^{\mbox{eff}})$
and $|C_7^{\mbox{eff}}|^2$. Of these, $C_{10}$ does not renormalize, 
however, there is a residual dependence on $\mu$ from the normalization 
for which inclusive semileptonic branching ratio is used, bringing in an 
extra $\alpha_s(\mu)$-dependence.

The scale-dependence of the individual contributions from the specified
Wilson coefficients to 
the branching ratios $\Delta \bar{\cal{B}}_s$  and $\Delta \bar{\cal{B}}_d$ 
and the branching ratios themselves, are shown in Fig.~1(a) and 
1(b), respectively.
 We find for the scale dependence of $\Delta \bar{{\cal{B}}}_s$ 
an uncertainty  
$(+9.0,-7.3) \%$, measured from the reference value $\mu=m_b$. This is to be
compared with the corresponding uncertainties  
$(+4.1,-1.3) \%$  calculated in the naive approach. The estimated 
$\mu$-dependent uncertainty in $\Delta \bar{{\cal{B}}}_d$
is found to be $(+7.7,-7.6) \%$, compared to $ 2 \%$ in the naive 
approach.

\begin{table}[h]
        \begin{center}
        \begin{tabular}{|c|c|c|c|c|}
        \hline
  \multicolumn{1}{|c|}{ \mbox{} }      &
  \multicolumn{1}{|c|}{ $\langle \Delta {\cal{B}}_s \rangle [ 10^{-6} ] 
$}   &
        \multicolumn{1}{|c|}{ $(a_{CP})_s [ \% ] $ }       &
  \multicolumn{1}{|c|}{ $\langle \Delta {\cal{B}}_d \rangle [ 10^{-8} ] $ 
}  &
        \multicolumn{1}{|c|}{ $(a_{CP})_d [ \% ] $  }       \\
        \hline
     AMM    & 2.22 & -0.19 & 9.61 &  4.40  \\
     KS     & 2.05 & -0.18 & 8.83 &  4.09  \\
     LSW    & 2.31 & -0.19 & 9.98 &  4.51  \\
    HQET    & 2.06 & -0.17 & 8.93 & 4.02  \\
 \hline
     $m_b=4.6$GeV & 2.15& -0.19 & 9.29 & 4.48  \\
     $m_b=5.0$GeV & 2.32& -0.18 & 10.03 & 4.29  \\
     $m_t=178.2$GeV & 2.36& -0.18 & 10.18 & 4.18  \\
     $m_t=168.2$GeV & 2.10& -0.20 & 9.06 &  4.63  \\
     $\Lambda_{QCD}^{(5)}=0.298$GeV & 2.20& -0.16 & 9.52 & 3.74  \\
     $\Lambda_{QCD}^{(5)}=0.157$GeV & 2.24& -0.22 & 9.70 & 5.03  \\
        \hline
        \end{tabular}
        \end{center}
\caption{ \it Values of the charge-conjugate averaged partial branching 
ratios
$\langle \Delta {\cal{B}}_s \rangle$ and $\langle \Delta {\cal{B}}_d
\rangle$ and the CP asymmetries $(a_{CP})_s$ and $(a_{CP})_d$, in the four
LD-approaches AMM \cite{amm91}, KS \cite{KS96}, LSW \cite{LSW97} 
and HQET \cite{buchallaisidorirey}, discussed in the text. In 
the top part of the table (above the horizontal line), the parameters
are fixed to their central values given in Table 1 and 
Eq.~(\ref{eq:ckmvalues}). In the lower part of
the table, the parametric dependence of the observables on $m_b$, $m_t$
and $\Lambda_{QCD}^{(5)}$, calculated using the AMM-approach, is listed.}
\label{tab:delB}
\end{table}

  The dependence of the 
charge-conjugate averaged 
branching ratios $\langle \Delta {\cal{B}}_s \rangle$ and
$\langle \Delta {\cal{B}}_d \rangle$, 
and the CP asymmetries $(a_{CP})_s$ and $(a_{CP})_d$ 
 on the four schemes concerning the $c\bar{c}$-contribution
is shown in the upper part of Table 2. For all these entries, we have
fixed the parameters to their central values given in Table 1 and
Eq.~(\ref{eq:ckmvalues}). The dependence of these observables 
on $m_b$, $m_t$ and
$\Lambda_{QCD}^{(5)}$, obtained in the AMM-scheme
by varying only one parameter at a time,
is shown in the lower part of Table 2.
For the central values of $\rho$ and $\eta$,
the partial branching ratios are found to
vary in the four approaches in the range: $ 2.05 \times 10^{-6} \leq \langle 
\Delta {\cal B}_s \rangle\leq 2.31 \times 10^{-6}$ and 
$ 8.83 \times 10^{-8} \leq \langle
\Delta {\cal B}_d \rangle\leq 9.98 \times 10^{-8}$. For the same values of
$\rho$ and $\eta$ but taking into account in addition 
the rest of the parametric uncertainties in Table 2, ${\cal B}_{sl}$, and the 
scale-dependence from Fig.~1(a) and 1(b), we find:
 \begin{eqnarray}
\langle \Delta {\cal{B}}_s \rangle &=& (2.22^{+0.29}_{-0.30}) \times 
10^{-6}~, \nonumber\\ 
\langle \Delta {\cal{B}}_d \rangle &=& (9.61^{+1.32}_{-1.47}) \times 
10^{-8}~.
\label{deltaerrors}
\end{eqnarray}
Thus, apart from the CKM-parametric dependence, we estimate $\pm 13\%$
uncertainty on $\langle \Delta {\cal{B}}_s \rangle$ and somewhat larger,
$\pm 15\%$, on $\langle \Delta {\cal{B}}_d \rangle$.
These errors are significantly larger than what one comes
across in the literature.  The present experimental bound is
${\cal B}(B \to X_s \ell^+ \ell^-) < 4.2 \times 10^{-5}$ (at 90\% C.L.)
\cite{cleobsll97}. We are not aware of a corresponding bound on ${\cal
B}(B \to X_d\ell^+ \ell^-)$. 
 
\begin{figure}[htb]
\vskip -0.2truein
\centerline{\epsfysize=3.5in
{\epsffile{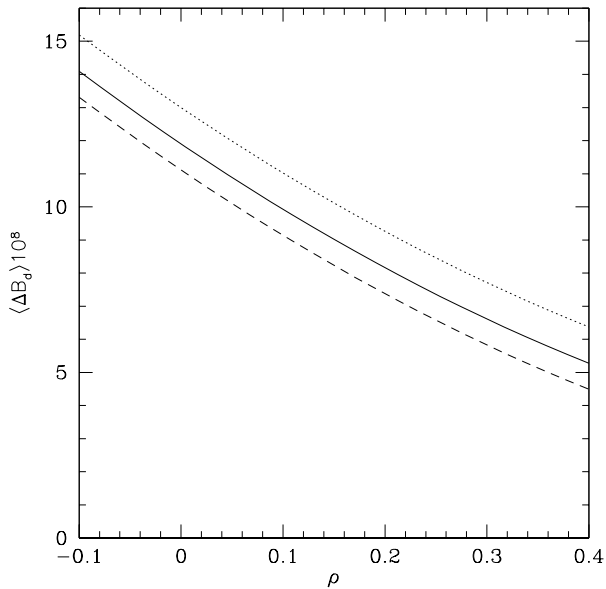}}}
\vskip -0.3truein
\caption[]{ \it The charge-conjugate averaged partial branching ratio
$\langle \Delta {\cal{B}}_d \rangle$ in the HQET-approach for 
the decay $B \to X_d 
\ell^+ \ell^-$ as a function of the CKM parameter $\rho$ for three values
of $\eta$; solid curve ($\eta=0.383)$, dotted curve $(\eta =0.5)$,
dashed curve ($\eta=0.27)$. }
 \label{fig:rhodep}
\end{figure}

The branching ratio $\langle \Delta {\cal{B}}_d 
\rangle$, calculated in HQET, is shown in Fig.~\ref{fig:rhodep}
as a function of the CKM parameter $\rho$ for three fixed values of 
$\eta$, which correspond to the central value and the
$95\%$ C.L. bounds given in Eq.~(\ref{eq:ckmvalues}).
The other input parameters have been fixed to their central values given in
Table~\ref{parameters}. In the allowed CKM parameter space,  
this partial branching ratio
varies by a factor 3. As the theoretical error 
from the rest of the parameters is estimated to be $\pm 15\%$, the
measurement of  $\langle \Delta {\cal{B}}_d \rangle$ should 
allow to determine $\rho$ 
and $\eta$. The ratio $\Delta {\cal R}=\langle \Delta {\cal{B}}_d 
\rangle/\langle \Delta {\cal{B}}_s \rangle$ has lot less theoretical error,
as shown below.

\begin{figure}[htb]
\vskip -0.2truein
\centerline{\epsfysize=3.5in
{\epsffile{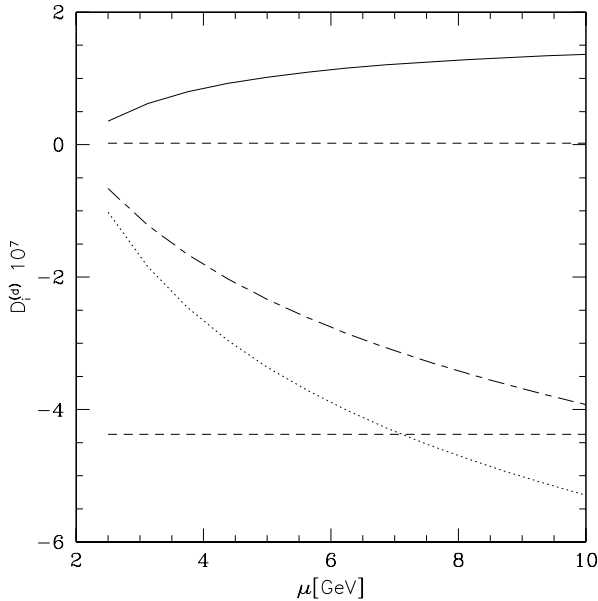}}}
\vskip -0.3truein
\caption[]{ \it 
Renormalization scale $(\mu)$-dependence of the individual contributions 
and the
function $D_i^{(d)}$, calculated in the AMM-approach.
The solid and dotted curves correspond to the 
contributions proportional to the effective Wilson coefficients  
$C_7^{\mbox{eff}} Im(C_9^{\mbox{eff}}|_u)$ and
$Im(C_9^{\mbox{eff}}|_u C_9^{\mbox{eff} \ast}|_t)$, respectively. 
The naive $\mu$ dependence is shown by the long-short dashed curve.
The resulting $\mu$
uncertainty in the independent approach is bounded by the dashed lines.} 
\label{fig:Dimudep}
\end{figure}

The CP asymmetry, $(a_{CP})_s$ defined in eq.~(\ref{eq:acp}) in the $b 
\to s$ case in the SM is small.
Hence its measurement can be used to search for new sources of CP 
violation in the $b \to s \ell^+ \ell^-$ transition.
Numerically, the CP asymmetries are more uncertain reflecting
in particular the scale-dependence of the functions $D^{(q)}_i$.
A qualitatively similar behaviour has also been noted for
the CP asymmetries in the radiative decays $B \to X_s + \gamma$ and $B 
\to X_d + \gamma$ in \cite{AAG98}. However, the scale-dependence of the
CP asymmetries is more marked in the decays
$B \to (X_s,X_d) \ell^+ \ell^-$ due to cancellations in two
different products of the Wilson coefficients entering in $D^{(q)}_i$.
(Specifically, between $ C_7^{\mbox{eff}} Im(C_{9q}^{\mbox{eff}}|_u)$ and 
$Im(C_{9q}^{\mbox{eff}}|_u C_{9q}^{\mbox{eff} \ast}|_t)$, with 
$C_{9q}^{\mbox{eff}}|_x$ denoting the part in $C_{9q}^{\mbox{eff}}$ 
which is proportional to the CKM factor $\lambda_x^{(q)}$.)
This can be seen in Fig.~\ref{fig:Dimudep}, where
we show the $\mu$-dependence of the two mentioned contributions in
$D_i^{(d)}$, and the function $D_i^{(d)}$ itself calculated in the naive
and independent approaches. The function $D_i^{(s)}$ is very similar and
hence not shown.
 The $\mu$-dependence of $D_i^{(d)}$ in the naive approach,
shown by the long-short dashed curve, is very
marked and it gets further accentuated in the independent approach, shown
by the two dashed curves.
For the central values of 
the CKM parameters and estimating the $\mu$-dependence in the 
independent approach, we find:
 \begin{eqnarray}
(a_{CP})_s &=&-(0.19^{+0.17}_{-0.19}) \%~, \nonumber\\
(a_{CP})_d&=&(4.40^{+3.87}_{-4.46}) \% .
\label{deltacp}
\end{eqnarray}
 The corresponding numbers in the naive scale-dependent method are:
$(a_{CP})_s =-(0.19^{+0.12}_{-0.13}) \% $, and
$(a_{CP})_d=(4.40^{+2.77}_{-3.23}) \%$.
In either case, Fig.~\ref{fig:Dimudep} underscores the importance of 
calculating the
next-to-leading order effects in $(a_{CP})_q$.


\subsection{Extraction of $\left|{V_{td} \over V_{ts}}\right|$ }

\begin{figure}[htb]
\vskip -0.2truein
\centerline{\epsfysize=3.5in
{\epsffile{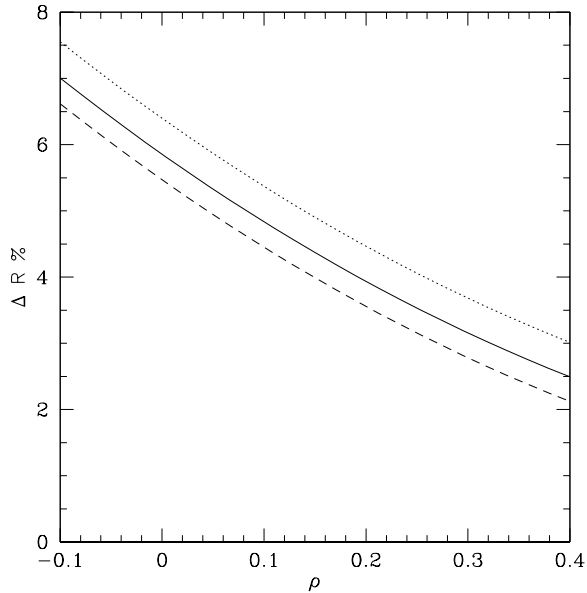}}}
\vskip -0.3truein
\caption[]{ \it The ratio $\Delta {\cal{R}}$ defined in Eq.~(\ref{eq:R}),
calculated in the HQET-approach, 
as a function of $\rho$ for three values
of $\eta$; solid curve ($\eta=0.383)$, dotted curve ($\eta=0.5)$,
dashed curve $(\eta=0.27)$. } \label{fig:raterhodep}
\end{figure}

For a precise determination of 
$\frac{|V_{td}|}{|V_{ts}|}$ (equivalently the CKM parameters), we calculate 
the ratio:
 \begin{eqnarray}
\label{eq:R}
\Delta {\cal{R}}\equiv \frac{\langle \Delta {\cal{B}}_d \rangle }
{\langle \Delta {\cal{B}}_s \rangle } \; .
\end{eqnarray}
In terms of the CKM parameters and the functions 
$D_t^{(s)}$ and $D_j^{(d)}$ with $j=t,u,r$, defined earlier:
 \begin{eqnarray}
\label{eq:delr}
\Delta {\cal{R}} =
\lambda^2 \frac{  (1-\bar{\rho})^2+\bar{\eta}^2 ) D_t^{(d)}
+( \bar{\rho}^2+\bar{\eta}^2 ) D_u^{(d)} + ( \bar{\rho} (1-\bar{\rho}) -
 \bar{\eta}^2 ) D_r^{(d)}}{ ( 1-\lambda^2 (1-2 \rho) ) D_t^{(s)} } \; ,
\end{eqnarray}
where we have neglected terms proportional to  
$\lambda_u^{(s)}/\lambda_t^{(s)}$. 
A simpler form for $\Delta {\cal{R}}$ follows, if one notes that the 
functions $D_t^{(d)}$ and $D_t^{(s)}$ are equal for all practical purposes
(see Table 3). Hence, setting $D_t^{(d)}=D_t^{(s)}$, one has
\begin{equation}
\label{deltaR}
\Delta {\cal{R}} = \lambda^2\frac{  (1-\bar{\rho})^2+\bar{\eta}^2 )}
{ ( 1-\lambda^2 (1-2 \rho)} \left[ 1 + \frac{( \bar{\rho}^2+\bar{\eta}^2 )}
{(1-\bar{\rho})^2+\bar{\eta}^2 )} \frac{D_u^{(d)}}{D_t^{(s)}} + \frac{( 
\bar{\rho} 
(1-\bar{\rho}) - \bar{\eta}^2 )}{(1-\bar{\rho})^2+\bar{\eta}^2 )}
\frac{ D_r^{(d)}}{D_t^{(s)}}
\right] ~.
\end{equation}
The overall CKM factor is just the ratio $\vert V_{td}\vert^2/\vert 
V_{ts}\vert^2$.
Note that the first (and dominant) term is independent of the dynamical
details. The ratio $D_u^{(d)}/D_t^{(s)}$ is found to be numerically small 
(but model dependent, varying between $1.03 
\times 10^{-2}$ for the KS-approach and $2.16 \times 10^{-2}$ for the LSW 
approach). The ratio $D_r^{(d)}/D_t^{(s)}$ is, in general, larger and it 
depends more sensitively on the estimate of the 
long-distance $c\bar{c}$-contribution,
varying between $+0.14$ (for the LSW-approach) and $-0.12$ (in HQET).
However, the multiplicative CKM factor accompanying this term
in Eq.~(\ref{deltaR}) being small comes to rescue. For example, for
$\bar{\rho}=0.151$ and 
$\bar{\eta}=0.374$, this factor is only $-0.012$. Hence, for these 
values, we find $\Delta {\cal{R}} = (4.32 \pm 0.03) \%$. For other values
of the CKM parameters, the  uncertainty is larger and we
quantify it later.
The ratio $\Delta {\cal{R}}$ as a function of $\rho$ is shown 
in Fig.~\ref{fig:raterhodep} for the HQET-method. The three curves
correspond to $\eta=0.5$ (dotted curve), $\eta=0.383$ (solid curve),
and $\eta =0.27$ (dashed curve).

\begin{table}[h]
        \begin{center}
        \begin{tabular}{|c|c|c|c|c|c|c|}
        \hline
        \multicolumn{1}{|c|}{ \mbox{} }      &
        \multicolumn{1}{|c|}{ $D_t^{(d)} [ 10^{-6} ] $}       &
        \multicolumn{1}{|c|}{ $D_u^{(d)} [ 10^{-8} ] $ }       & 
        \multicolumn{1}{|c|}{ $D_t^{(s)} [ 10^{-6} ] $ }       &
        \multicolumn{1}{|c|}{ $D_r^{(d)} [ 10^{-8} ] $  }      &
        \multicolumn{1}{|c|}{ $D_i^{(d)} [ 10^{-7} ] $ }       &
        \multicolumn{1}{|c|}{ $D_i^{(s)} [ 10^{-7} ] $  }       \\
        \hline 
     AMM    & 2.31 & 3.75 & 2.30 &  20.96 &-2.34 & -2.34 \\
     KS     & 2.12 & 2.18 & 2.11 &   1.42 &-2.00 & -2.05\\
     LSW    & 2.40 & 5.16 & 2.39 &  32.59 &-2.50 & -2.43\\   
     HQET    & 2.14 & 2.88 & 2.13 & -24.89&-1.99 & -1.94 \\
 \hline
     $m_b=4.6$GeV & 2.24& 4.48 & 2.22 & 26.83   &-2.31&-2.26  \\
     $m_b=5.0$GeV & 2.41& 3.47 & 2.40 & 18.86   &-2.39&-2.31 \\
     $m_t=178.2$GeV & 2.45& 3.75 & 2.44 & 21.89 &-2.36&-2.35  \\
     $m_t=168.2$GeV & 2.18& 3.75 & 2.17 & 21.61 &-2.33&-2.33  \\
     $\Lambda_{QCD}^{(5)}=0.298$GeV & 2.29& 3.39 & 2.28 & 20.71 &-1.97&-1.95 \\
     $\Lambda_{QCD}^{(5)}=0.157$GeV & 2.33& 4.15 & 2.32 & 21.35 &-2.70&-2.73 \\
        \hline
        \end{tabular}
        \end{center}
\caption{ \it Values of the functions $D_j^{(d)}$, $j=u,t,r,i$ and
$D_t^{(s)},D_i^{(s)}$ defined in 
eq.~(\ref{eq:delr}) and (\ref{eq:deli})
in the four schemes discussed in the text for the
central values of the input parameters. The entries
below the horizontal line correspond to using the AMM scheme, and 
varying the input parameters, one each at a time, fixing the rest to
their central values.}
\label{tab:Di}
\end{table}

\begin{figure}[htb]
\vskip -0.2truein
\centerline{\epsfysize=3.5in
{\epsffile{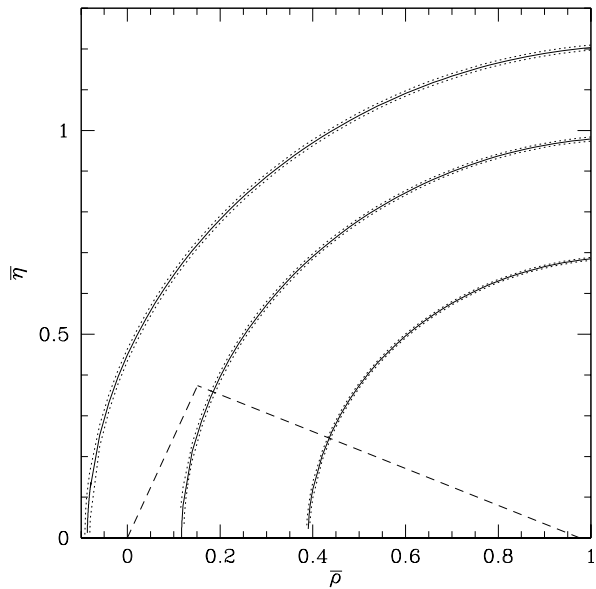}}}
\vskip -0.3truein
\caption[]{ \it Contours in the $(\bar{\rho},\bar{\eta})$ plane
following from  assumed values of the ratio 
$\Delta {\cal{R}}$; outer curve ($\Delta {\cal{R}}=0.06$),
central curve  ($\Delta {\cal{R}}=0.04$), inner curve 
 ($\Delta {\cal{R}}=0.02$). 
The overlapping curves for each value of $\Delta {\cal{R}}$ represent 
the uncertainty due to the renormalization scale.
Also shown is the unitarity triangle corresponding to the central values 
of the CKM parameters from the analysis of Ref.~\cite{smele98}.}
\label{fig:sduncert}
\end{figure} 
\begin{figure}[htb]
\vskip -0.2truein
\centerline{\epsfysize=3.5in
{\epsffile{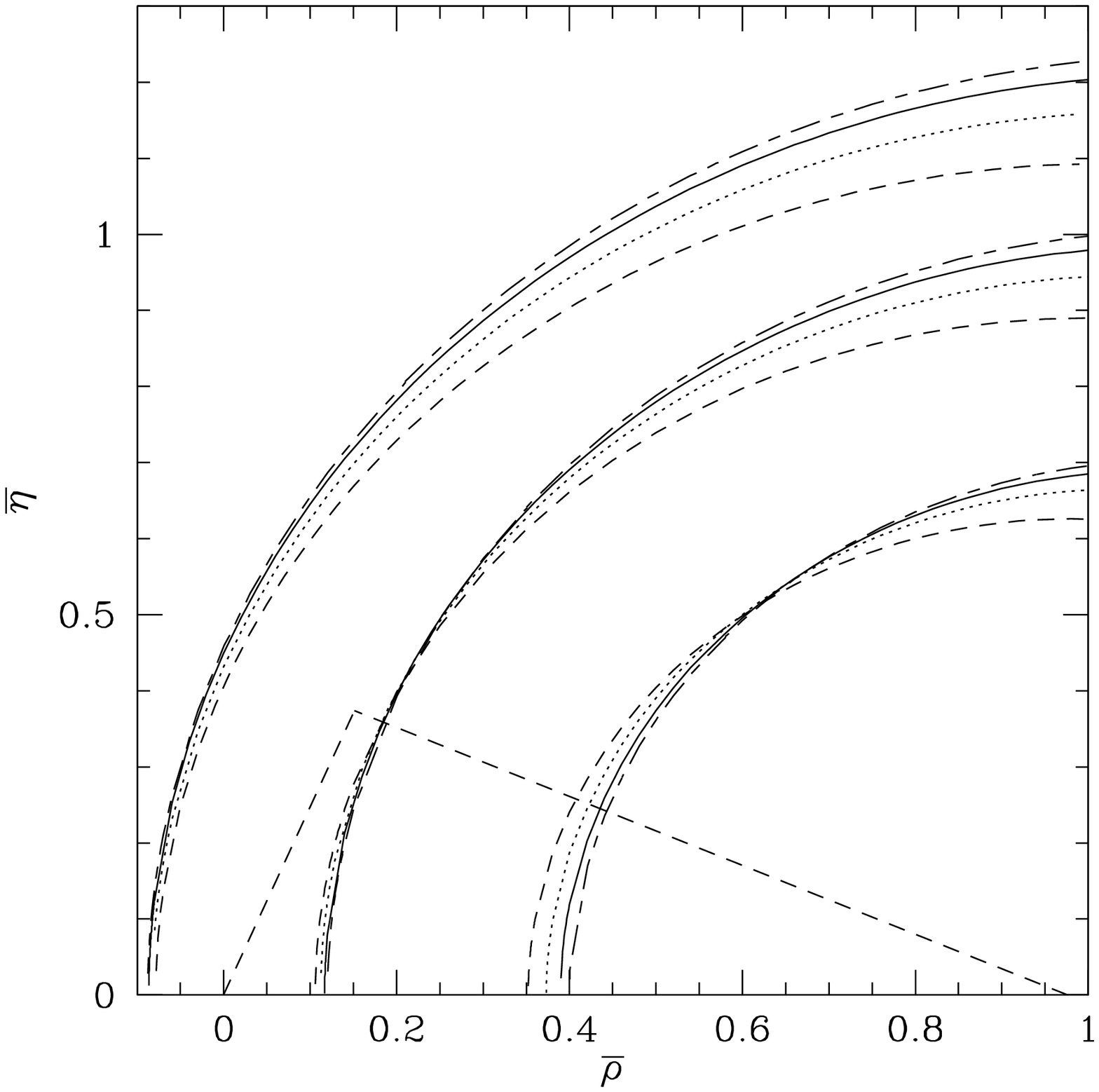}}}
\vskip -0.3truein
\caption[]{ \it
Contours in the $(\bar{\rho},\bar{\eta})$ plane
following from  assumed values of the ratio
$\Delta {\cal{R}}$; outer curve ($\Delta {\cal{R}}=0.06$),
central curve  ($\Delta {\cal{R}}=0.04$), inner curve
 ($\Delta {\cal{R}}=0.02$). The solid, dotted, dashed, long-short dashed 
lines correspond to
the AMM, KS, HQET and LSW approaches, respectively, 
for the central values of the parameters given in Table 1.
Also shown is the unitarity triangle corresponding to the central values of 
the CKM parameters from the analysis of ref. \cite{smele98}.}
\label{fig:lduncert}
\end{figure} 
We now evaluate the theoretical precision in the determination of 
$\left|{V_{td} \over V_{ts}}\right|$ from an eventual measurement of
$\Delta {\cal R}$. The other uncertainties being insignificant, there are
basically two sources of errors: (i) a small residual
scale-dependence, and (ii) the LD-scheme-dependent uncertainty, 
which depends on the parameters $\rho$ and $\eta$.
 In Fig.~\ref{fig:sduncert} we show the 
constraints 
on $\rho$ and $\eta$ from an assumed value of $\Delta {\cal R}$ with the
LD-effects calculated in the AMM-approach.
For each value of $\Delta {\cal R}$, the practically overlapping curves
represent the effect of varying $\mu$ in the range $m_b/2 \leq \mu \leq
2 m_b$. Numerically, the net $\mu$ 
uncertainty on the ratio $\Delta \cal{R}$ is found to be $\pm 0.6 \%$.
The effect of the errors of $m_t, \alpha_s(m_Z)$ and the $b$-quark mass are
smaller and not shown.

The potentially largest uncertainty in $\Delta \cal{R}$, due to the 
LD-effects, is shown in Fig.~\ref{fig:lduncert}, where we have plotted the
constraints on $\rho$ and $\eta$ from  assumed values of $\Delta \cal{R}$.
The four curves shown correspond to the LD-schemes: AMM, KS, HQET and LSW.
As remarked earlier, the LD-related uncertainty is
vanishingly small for the central values of $\rho$ and $\eta$, i.e. at or
close to the apex of 
the drawn triangle. However, for other points in the $(\rho,\eta)$-plane, 
the uncertainty is perceptible but still small, except for regions of the
$(\rho,\eta)$-plane which are already ruled out from the existing CKM fits. 

\section{Theoretical Precision on $\vert V_{td}/V_{ts}\vert$
from $B$ Decays}

 The ratio $\Delta {\cal R}$ should be measurable
at the Tevatron, the later phase of the B-factories, and certainly
at the LHC. 
The merit of 
$\Delta {\cal R}$ lies in the theoretical precision
on $\vert V_{td}/V_{ts}\vert$ (or on the unitarity triangle) which we have 
estimated
here and found to be quite competitive with other proposals in the market,
some of which are reviewed below.

The $B^0$-$\overline{B^0}$ mixing ratio $\Delta 
M_s/\Delta M_d$ can be expressed as follows:
\begin{equation}
\frac{\Delta M_s}{\Delta M_d}=\frac{M_{B_s}}{M_{B_d}}
 \frac{(f_{B_s}^2 \hat{B}_{B_s})}{(f_{B_d}^2 \hat{B}_{B_d})}
\vert \frac{V_{ts}}{V_{td}}\vert^2~.
\end{equation}
The achievable accuracy on $V_{td}/V_{ts}$ depends, apart from the
experimental measurement error, on the
knowledge of the ratio of the hadronic matrix elements $\xi \equiv
f_{B_d}\sqrt{B_{B_d}}/f_{B_s}\sqrt{B_{B_s}}$, for which the current 
Lattice
estimate is $\xi=1.14 \pm 0.06 \pm 0.03 \pm 0.10$ \cite{Draper98}.
The errors reflect, respectively, the actual calculational error of
this ratio in the quenched approximation, estimated effects of unquenching,
and from chiral loops.  Thus, the present theoretical error on this
quantity is of $O(10\%)$ and it remains a theoretical challenge to
improve this significantly. However, the measurement
of $\Delta M_s$, for which the present experimental lower bound is 12.4
ps$^{-1}$ (at 95\% C.L.)\cite{parodi98}, may turn out to provide the first
measurement of
$V_{td}/V_{ts}$, as the central value of $\Delta M_s$ in the SM is around 
$14~\mbox{ps}^{-1}$ \cite{smele98,parodi98,alilondon98}, which is 
not too far from the present limit.

Theoretical precision on $\Delta {\cal R}$ is comparable to the
one on the corresponding ratio of the branching ratios involving the
CKM-suppressed decay $B \to X_d + \gamma$ and the
CKM-allowed decay  $B \to X_s + \gamma$ \cite{AG92,AAG98}.
Defining the ratio of the branching ratios as (implied are 
charge-conjugate averages)
\begin{equation}
R(d\gamma)/s\gamma) \equiv \frac{\langle {\cal B}(B \to X_d + \gamma)
\rangle}{\langle {\cal B}(B \to X_s + \gamma)\rangle},
\end{equation}
the ratio $R(d\gamma)/s\gamma)$ gives a constraint on the CKM
matrix elements which is very similar to the one given by 
$\Delta {\cal R}$ (compare Eq.~(26) in Ref.~\cite{AAG98} and 
Eq.~(\ref{deltaR}) here). 
Theoretical error on $R(d\gamma/s\gamma)$ is estimated to be at most a few 
percent in \cite{AAG98}, comparable to the one on $\Delta {\cal R}$.
In hadronic collisions, the ratio
$\Delta {\cal R}$ is more likely to be measured than $R(d\gamma/s\gamma)$. 

 We also mention here the exclusive radiative decays $B \to (\rho,\omega)
\gamma$
and $B \to K^* \gamma$, whose ratios of the branching ratios can also be
used to determine $\vert V_{td}/V_{ts}\vert$ \cite{ABS94}.
 The expected theoretical accuracy on the
ratio ${\cal B}(B^\pm \to \rho^\pm + \gamma)/{\cal B}(B^\pm \to K^{*\pm}+ 
\gamma)$
is, however, not anticipated to be better than $O(20\%)$  \cite{AB95}.
The corresponding LD-corrections in the ratios of neutral $B$-decays,
${\cal B}(B^0 \to (\rho^0, \omega) + \gamma)/{\cal B}(B^0 \to K^{*0}
+ \gamma)$ are
expected to be smaller \cite{AB95,Donoghue97} due to their being both 
colour and (electric)-charge suppressed, hence reducing the theoretical 
uncertainty, but probably  not better than $\pm 10\%$. 
Finally, we also note
the constraints on $\vert V_{td}/V_{ts}\vert$, which can be obtained from the
measurements of the ratios of some exclusive two-body non-leptonic decays, 
such as
${\cal B}(B^0 \to \overline{K^*} K^0)/{\cal B}(B^0 \to \phi K^0)$, advocated
in Ref.~\cite{GR96}. This method may provide interesting results on
the CKM ratio, but once data are available on the FCNC radiative and 
semileptonic decays discussed above, they are expected to provide 
more reliable information on the CKM matrix elements 
$V_{td}$ and $V_{ts}$. In particular, the ratio $\Delta {\cal R}$ may
provide one of the most precise determinations of $\vert V_{td}/
V_{ts}\vert$. 

  We hope that the results presented here will help focus 
attention on experimental measurements of the branching ratios and 
CP asymmetries in the FCNC decays $B \to (X_d,X_s) \ell^+ \ell^-$. We
also underline the need to calculate the next-to-leading order corrections
in the CP asymmetries to tame the scale dependence.


\bigskip
\noindent
{\Large \bf Acknowledgements}

G.~H. would like to thank Gino Isidori and Frank Zimmermann for helpful 
discussions and the Theoretical Physics groups at CERN and SLAC for the 
hospitality during her stay, where part of the work reported here has been 
done. She gratefully acknowledges a fellowship from the European 
Community under contract number FMRX-CT98-0169.


\begin{thebibliography}{1}
\bibitem{CKM} N. Cabibbo, Phys.~Rev.~Lett. 10, 531 (1963); M. Kobayashi 
and K. Maskawa, Prog.~Theor.~Phys. 49, 652 (1973).

\bibitem{CDFVtb} M. Narain, in Proc.~of the Seventh International 
Symposium on Heavy Flavors, Santa Barbara, Calif., 1997.

\bibitem{Draper98} T. Draper, hep-lat/9810065, to be published in the
Proc. of the Conference Lattice '98, Boulder, Colorado, 1998.

\bibitem{E787} S. Adler et al. (E787 Collaboration), Phys.~Rev.~Lett. 79,
2204 (1997).

\bibitem{Ali97} A. Ali, report DESY 97-256, hep-ph/9801270; to be published
in Proc.~of the Seventh International Symposium on Heavy Flavors, Santa 
Barbara, Calif., 1997.
 
\bibitem{CLEO98}%
M.S.~Alam et al. (CLEO Collaboration), Phys.~Rev.~Lett.~74, 2885 (1995); 
 T. Skwarnicki (CLEO Collaboration), in Proc. of 
the conference ICHEP'98, Vancouver, 1998.

\bibitem{ALEPH98} R. Barate et al. (ALEPH Collaboration), Phys.~Lett. 
B429, 169 (1998).

\bibitem{PDG98} C. Caso et al. (Particle Data Group), Eur. Phys. J. 
C3, 1 (1998).

\bibitem{AGM93} A. Ali, C. Greub and T. Mannel, report DESY 93-016, ZU-TH 
4/93, IKDA 93/5, published in the Proc. of the ECFA Workshop on a 
European B Meson Factory, Eds. R. Aleksan and A. Ali, Hamburg 1993.  


\bibitem{buchallaisidorirey} G. Buchalla, G. Isidori and S.~-J.~Rey,
          Nucl.~Phys.~B511, 594 (1998).

\bibitem{Burasetal93} G. Buchalla and A.J. Buras, Nucl.~Phys. B400, 225 
(1993).

\bibitem{ALEPHbsnunu} ALEPH Collaboration, Contributed paper (PA10-019)
to the 28th. International Conference on High Energy Physics, Warsaw,
1996.

\bibitem{Zerwas98}
We thank Peter Zerwas for bringing this to our attention. See also, the
Proceedings of the Workshop on Physics and Detectors for a Linear $e^+ e^-$ 
Collider, Frascati, November 1998.

\bibitem{KMS} C.S. Kim, T. Morozumi and A.I. Sanda,
           Phys.~Rev.~D56, 7240 (1997).
%
\bibitem{AHHM97} A. Ali, L. T. Handoko, G. Hiller and T. Morozumi,
Phys.~Rev.~D55, 4105 (1997).
%
%
\bibitem{cslim89}
C.S.~Lim, T.~Morozumi and A.I.~Sanda, Phys.~Lett.~B218, 343 (1989);
N.G.~Deshpande, J.~Trampetic and K.~Panose, Phys.~Lett.~ B214, 467 (1988)
Phys.~Rev.~D39, 1461 (1989); P.J. O'Donnel and H.K.K.~Tung,
Phys.~Rev.~D43, 2067 (1991).
%
%
\bibitem{amm91}
      A. Ali, T. Mannel and T. Morozumi, Phys.~Lett.~B273, 505 (1991).


\bibitem{aliqcd}
        A. Ali and E. Pietarinen, Nucl.~Phys.~B154, 519 (1979); \\
        G. Altarelli et al., Nucl.~Phys.~ B208, 365 (1982).
%

\bibitem{AH98-12}
A.~Ali and G.~Hiller, Phys.~Rev. D58, 074001 (1998); D58, 071501 (1998).

\bibitem{AH98-3}
A.~Ali and G.~Hiller, preprint DESY 98-031, hep-ph/9807418 (submitted to
Phys. Rev. D).

\bibitem{MKS98}
D.~Melikhov, N.~Nikitin and S.~Simula, Phys.~Rev.~D57, 6814 (1998);
preprint hep-ph/9803343.
%
%
\bibitem{KS96}
F.~Kr\"uger and L.M.~Sehgal, Phys.~Lett.~B380, 199 (1996).
%
\bibitem{LSW97}
Z.~Ligeti, I.W.~Stewart and M.B.~Wise, Phys.~Lett.~ B420, 359 (1998).
%
\bibitem{Wolfenstein} L. Wolfenstein, Phys. Rev. Lett.~51, 1845 (1983).

\bibitem{KN98} A.L.~Kagan and M.~Neubert, report
CERN-TH-98-99, hep-ph/9805303.
%
\bibitem{KSCP97}
F. Kr\"uger and L.M. Sehgal, Phys.~Rev.~D55, 2799 (1997).
%
%
\bibitem{burasmuenz} 
 A. J. Buras and M. M\"unz, Phys.~Rev.~D52, 186 (1995).
%
%
\bibitem{misiakE}
     M. Misiak, Nucl.~Phys.~B393, 23 (1993) [E.~B439, 461 (1995)].
%


\bibitem{jezkuhn} A. Czarnecki, M. Je\.{z}abek and J. H. K\"uhn,
Acta.~Phys.~Pol.~B20, 961 (1989); \\ 
M. Je\.{z}abek and J. H. K\"uhn, Nucl.~Phys.~B320, 20 (1989).
%
%
\bibitem{effhamburas}
  A. J. Buras et al., Nucl.~Phys.~B424, 374 (1994). 
%
\bibitem{BLO94}
A.J. Buras, M.E. Lautenbacher and G. Ostermaier, 
Phys.~Rev.~D50, 3433 (1994).
 
\bibitem{smele98} S.~Mele, preprint hep-ph/9808411.
%
\bibitem{parodi98} F. Parodi, P.Roudeau and A. Stocchi, preprint 
hep-ph/9802289, and contributed paper \# 586 to ICHEP '98, Vancouver, 
July 1998.
%
\bibitem{alilondon98} A. Ali and D. London, Nucl.~Phys.~B (Proc. Suppl.) 54A,
297 (1997), and to be published.
%
%
\bibitem{cleobsll97}
S.~Glenn et al. (CLEO Collaboration), Phys.~Rev.~Lett.~80, 2289 (1998).
%

\bibitem{AG92} A. Ali and C. Greub, Phys.~Lett. B287, 191 (1992).
%
\bibitem{AAG98} A. Ali, H. Asatrian and C. Greub, Phys.~Lett. B429, 87 
(1998).
%
\bibitem{ABS94} A.~Ali, V.~Braun, and H.~Simma, Z.~f.~Phys.~C63, 437 (1994).
%
\bibitem{AB95} 
A.~Khodjhamirian, G.~Stoll, and D.~Wyler, Phys..~Lett.~B358, 129 (1995);\\
A.~Ali and V.~Braun, Phys.~Lett.~B359, 223 (1995).
%
\bibitem{Donoghue97}
J.F.~Donoghue, E.~Golowich, and A.A. Petrov, Phys. Rev.~D55, 2657 (1997).
%
\bibitem{GR96} M.~Gronau and J.L.~Rosner, Phys.~Lett.~B376, 205 (1996).
%
\end{thebibliography}
\end{document}